\documentclass{article}
\usepackage{algorithm}
\usepackage{algpseudocode}
\usepackage{amsmath}
\usepackage{adjustbox}
\usepackage{tabularx,caption}

\usepackage[preprint]{neurips_2025}

\usepackage[utf8]{inputenc} 
\usepackage[T1]{fontenc}    
\usepackage{hyperref}       
\usepackage{url}            
\usepackage{booktabs}       
\usepackage{amsfonts}       
\usepackage{nicefrac}       
\usepackage{microtype}      
\usepackage{xcolor}         
\usepackage{wrapfig}
\usepackage{tabularx}
\usepackage{graphicx}
\usepackage{multirow}   
\usepackage{adjustbox} 
\usepackage{amsmath}
\usepackage{amssymb}
\usepackage{enumitem}
\usepackage{subcaption} 
\title{Seeing Beyond Words: MatVQA for Challenging Visual-Scientific Reasoning in Materials Science}

\author{%
  Sifan Wu \\
  University of Montreal \& MILA\\
  \texttt{sifan.wu@umontreal.ca} \\
  \And
  Huan Zhang \\
  University of Montreal \& MILA\\
  \texttt{huan.zhang@umontreal.ca} \\
  \AND
  Yizhan Li \\
  University of Montreal \& MILA \\
  \texttt{yizhan.li@umontreal.ca} \\
  \AND
   Farshid Effaty \\
  University of Montreal \& MILA \\
  \texttt{farshid.effaty@umontreal.ca} \\
  \And
  Amirreza Ataei \\
  Chemia Discovery Inc. \\
  \texttt{amirreza.ataei@chemiadiscovery.com} \\
  \And
  Bang Liu \\
  University of Montreal \& MILA \\
  \texttt{bang.liu@umontreal.ca} \\
}

\begin{document}

\maketitle

\begin{abstract}
The emergence of Multimodal Large Language Models (MLLMs) that integrate vision and language modalities has unlocked new potentials for scientific reasoning, outperforming prior benchmarks in both natural language and coding domains. Current materials science evaluation datasets such as MaScQA and SciQA remain largely text-based and fail to capture the visual and research-level analytic complexity required in materials discovery and design. We introduce MatVQA, a scalable benchmark specifically designed to address this gap. Generated via an automated pipeline, MArxivAgent, from recent materials literature, MatVQA features 1325 questions across four critical structure-property-performance (SPP) reasoning tasks. Uniquely, MatVQA employs an iterative process to eliminate textual shortcuts, compelling MLLMs to perform fine-grained, low-level visual analysis of material imagery (e.g., microscopy, diffraction patterns) integrated with multi-step scientific reasoning. Benchmarking 17 open- and closed-source MLLMs on MatVQA reveals substantial gaps in current multimodal reasoning capabilities. MatVQA benchmark data, along with evaluation code, is publicly available in \href{https://anonymous.4open.science/r/matvqa-1E01}{https://anonymous.4open.science/r/matvqa-1E01/README.md} to catalyze further research in applying MLLMs to complex materials science problems.
\end{abstract}

\section{Introduction}
Multimodal Large Language Models (MLLMs) have recently demonstrated remarkable success in a variety of applications, including natural language understanding~\cite{bedi2024testing, brown2020language, hollmann2025accurate}, code generation~\cite{jain2024livecodebench, nam2024using}, and diverse scientific domains~\cite{latif2024physicsassistant, song2023honeybee, zhang2024honeycomb, zhang2024mathverse}. Their ability to interpret and reason over complex, multimodal data while integrating specialist domain knowledge is pivotal for tackling challenging problems in fields such as mathematics, medicine, finance, and geospatial sciences. While MLLMs have shown promise in facilitating scientific discovery in areas like biology~\cite{jung2024llm, luu2024bioinspiredllm}, chemistry~\cite{jablonka202314, zhang2024chemllm}, software engineering~\cite{belzner2023large}, and healthcare~\cite{peng2023study, nazi2024large}, their application to research-level reasoning in highly specialized, multimodal scientific tasks, particularly within materials science, remains less explored.

Materials science, an interdisciplinary field at the intersection of physics, chemistry, and often biology, demands extensive domain knowledge and sophisticated reasoning capabilities to understand and design material. Initial efforts have applied Large Language Models (LLMs) to augment materials science research, such as the HoneyComb agent for material-related question answering~\cite{zhang2024honeycomb}. However, these approaches, along with existing materials science datasets like MaScQA~\cite{zaki2024mascqa}, SciQA~\cite{auer2023sciqa}, and MSE-MCQs~\cite{wang2024evaluating}, predominantly focus on text-based question answering. Consequently, they often overlook the critical aspect of visual perception and lack questions that require deep, research-level reasoning. The recent emergence of benchmarks like MicroVQA~\cite{burgess2025microvqa} for biological microscopy underscores the necessity of evaluating MLLMs on research-grade multimodal tasks. To fill this void, we introduce MatVQA. MatVQA is distinguished by its design for easy scalability and its specific focus on assessing two capabilities largely underaddressed by existing benchmarks: the complex, multi-step reasoning constructed from fundamental scientific principles, and the demanding, low-level visual perception required to interpret nuanced experimental data. 
\begin{wrapfigure}{r}{0.5\textwidth}        
  \vspace{-0.2em}
  \centering
  \includegraphics[width=\linewidth]{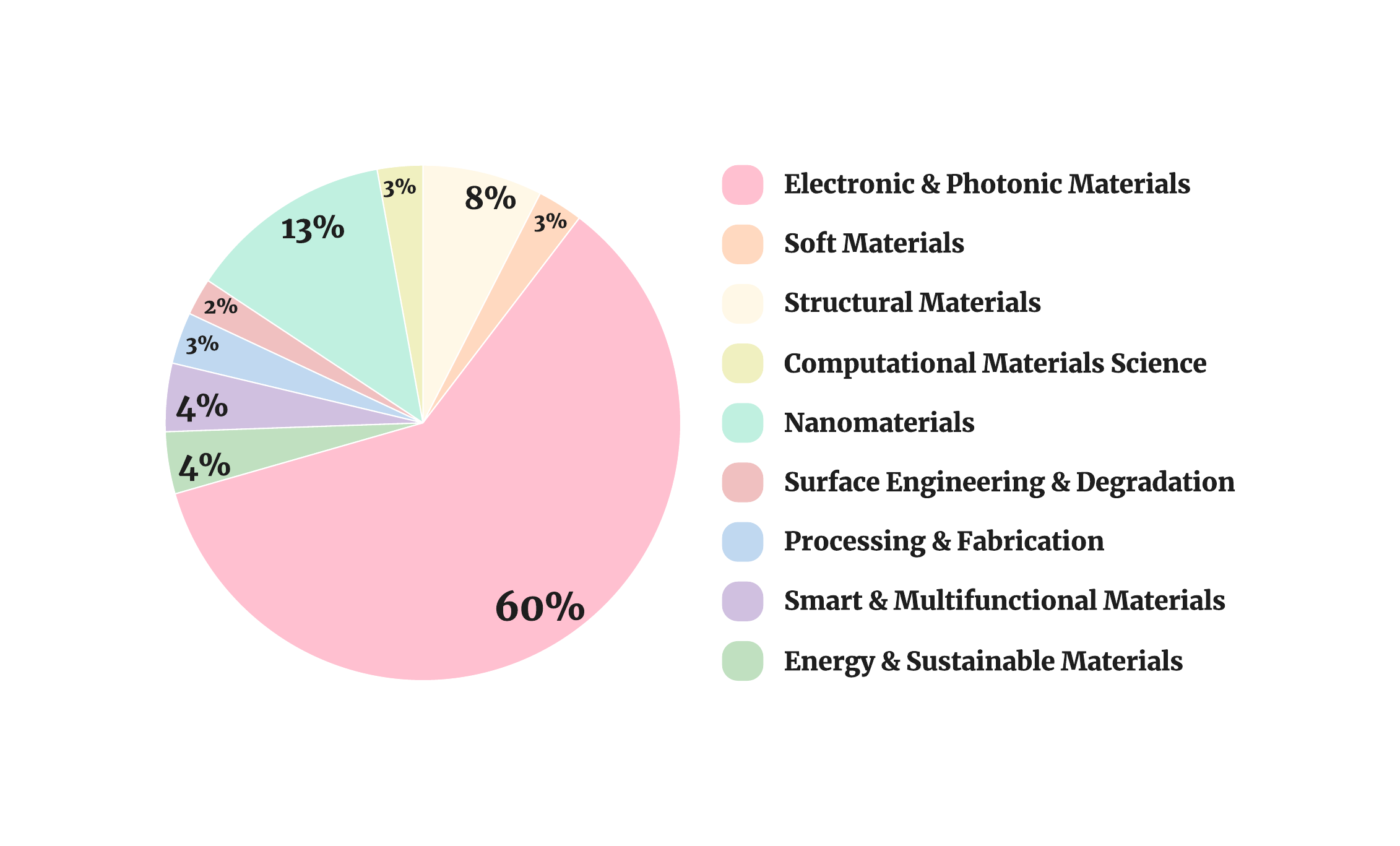}
  \caption{Domain attribution for \textbf{MatVQA}}
  \label{fig:domain-ratio}
  \vspace{-1.5em}
\end{wrapfigure}

MatVQA is built by \textbf{MArxivAgent}, an automated and verifiable pipeline engineered for the efficient generation of challenging multiple-choice questions (MCQs) directly from arXiv materials-science papers. This automated construction underpins MatVQA's inherent scalability, facilitating ongoing expansion and adaptation. After generation by advanced LLMs, a random 20\% of the MCQs are vetted by domain experts to ensure quality. Grounded in real-world literature, the benchmark targets four research-critical structure–property–performance (SPP) tasks—since they map how a material’s \emph{structure} governs its measurable \emph{properties} and ultimate \emph{performance}. These SPP tasks, derived from core components of scientific inquiry, by their very nature necessitate the complex reasoning MatVQA aims to evaluate, requiring tight integration of visual and textual evidence and collectively probing the core cognitive operations of materials research:
\begin{enumerate}[label=(\arabic*),nosep]
\item \textbf{Quantitative SPP}: Rigorously quantify how specific structural descriptors modulate properties and performance, providing the numerical design rules essential for rational materials engineering.
\item \textbf{Comparative SPP}: Systematically contrast competing structures under identical conditions to pinpoint the superior configuration, mirroring the materials-selection step in experimental practice.
 \item \textbf{Causal SPP}: Infer mechanistic cause-and-effect links between structure and performance, advancing understanding beyond correlation and enabling mechanism-driven optimisation.
 \item \textbf{Hypothetical variation}: Predict the ramifications of untested structural modifications, permitting in-silico exploration of the design space and accelerating discovery.
\end{enumerate}
These tasks mirror the questions materials scientists pose when characterising, designing, and optimising materials, and they rigorously test an MLLM’s ability to perform fine-grained visual perception, numerical reasoning, comparative analysis, causal inference, and forward prediction. As shown in Table~\ref{tab:comp}, MatVQA contains 1325 multiple-choice questions automatically generated by advanced language models and randomly verified by domain experts carefully.

Motivated by the forensic study of multimodal biases in \cite{burgess2025microvqa}, we identify two textual artifacts that can subvert genuine vision–language evaluation:  
\textbf{(i) Language shortcuts} appear when the prompt provides verbose image descriptions, weak distractors, or stylistic cues that allow the answer to be inferred from the text alone.  
\textbf{(ii) Caption shortcuts} happen when the information embedded in its stem or options—typically paraphrased from the figure caption—suffices to answer the question without inspecting the figure itself.  
Caption shortcuts are especially prevalent in materials-science corpora because captions explicitly condense the key morphological, crystallographic, or spectroscopic observations. Their inadvertent reuse limits MCQs to high-level descriptors (e.g., phase label, mean grain size) and excludes the low-level visual cues—diffraction peaks, defect textures, subtle contrast variations—crucial for authentic reasoning. Additional leakage stems from distractors generated with caption vocabulary, reliance on caption-embedded numerical values, and minimal emphasis on spatial or pixel-level patterns.  

Eliminating both shortcut classes is therefore imperative: it compels models to ground their answers in fine-grained visual evidence and provides a benchmark that more faithfully measures genuine vision–language competence.

To excise these artifacts, MArxivAgent executes an \emph{iterative shortcut-elimination loop}.  After initial question synthesis, an evaluator agent answers the MCQ using (a) only the stem and options and (b) the stem, options plus caption but \emph{without} the image.  Success in either mode triggers a rewriter that removes or rephrases the incriminating text while a consistency checker enforces fidelity to the original scientific claim.  By progressively eliminating both language and caption shortcuts, we elevate the benchmark from coarse- to fine-grained difficulty: solving the final questions requires precise, low-level visual scrutiny (e.g., counting diffraction spots, discerning lattice fringes) coupled with multi-hop scientific reasoning.  This refinement is essential for measuring the true multimodal competence that front-line materials research demands.

We evaluated a suite of 17 open- and closed-source MLLMs on MatVQA and compared the performance of a select subset against human experts and vision-language model baselines. In summary, our key contributions are:
\begin{itemize}[nosep,itemsep=1pt]
    \item We release \textbf{MatVQA}, the first benchmark designed to evaluate research-level multimodal reasoning in varies domain of materials science.
    \item We propose four Structure-Property-Performance (SPP) tasks that encapsulate core scientific inquiries regarding material structure, properties, and performance.
    \item We design \textbf{MArxivAgent}, a fully automated, three-stage pipeline that (i) extracts reasoning paths from scientific literature, (ii) iteratively eliminates language shortcuts, and (iii) subsequently removes caption shortcuts, producing high-difficulty, visually grounded MCQs.
\end{itemize}
This work aims to foster the development of MLLMs capable of contributing meaningfully to materials science research by providing a challenging and relevant evaluation standard.

\begin{table*}[t]
  \centering
  \caption{Overview of multimodal science benchmarks and detailed MatVQA attributes (in (b), SPP represents \textbf{structure–property–performance}).}
  \label{tab:comp}
  \begin{subtable}[t]{0.65\textwidth}  
    \centering
    \caption{Comparison with current multimodal benchmarks for Science}
    \label{tab:science_benchmarks}
    \small
    \begin{tabularx}{\linewidth}{@{}l*{4}{c}@{}}
      \toprule
      Benchmark & Level & Domain & Source & Size \\
      \midrule
      ScienceQA~\cite{lu2022learn}        & High-school & Science   & Exams     & 16.8k \\
      MicroBench~\cite{lozano2024micro}    & Graduate & Microscopy   & Datasets  & 17.2k \\
      MMSci~\cite{li2024mmsci}            & PhD      & Science      & Figures   & 7,132 \\
      \midrule 
      MacBench~\cite{alamparamacbench}     & Research & Chem\&Mat   & Lab       & 628 \\
      LabBench~\cite{laurent2024lab}       & Research & Biology      & WebQA     & 181 \\
      MicroVQA~\cite{burgess2025microvqa}  & Research & Microscopy   & Expert    & 1,042 \\
      \textbf{MatVQA} (Ours)                  & Research & Materials    & Paper     & 1,325 \\
      \bottomrule
    \end{tabularx}
  \end{subtable}\hfill
  \begin{subtable}[t]{0.33\textwidth}  
    \centering
    \caption{MatVQA benchmark attributes}
    \label{tab:matvqa_attributes}
    \small
    { 
    \renewcommand{\arraystretch}{0.88} 
    \begin{tabular}{@{}ll@{}}
      \toprule
      \textbf{MatVQA feature} & \textbf{Value} \\
      \midrule
      Total questions      & 1,325 \\
      Causal SPP questions &    950 \\
      Comp SPP questions   &    112 \\
      Hypo SPP questions   &    256 \\
      Quan SPP questions   &      7 \\
      \midrule
      Unique images        & 378 \\  
      Unique papers        & 44 \\  
      Research areas       & 32 \\
      \bottomrule
    \end{tabular}
    } 
  \end{subtable}
\end{table*}

\section{Related Work}
\subsection{MLLM Reasoning Benchmarks}
Recent work has introduced a range of benchmarks that probe how well multimodal large language models (MLLMs) integrate visual and textual reasoning. MATHVISTA~\cite{lu2024mathvistaevaluatingmathematicalreasoning} tests fine-grained visual understanding and compositional math reasoning with thousands of expert-designed problems that expose the gap between today’s models and human mathematicians. Similarly, benchmarks focusing on code reasoning, such as CRUXEval~\cite{gu2024cruxevalbenchmarkcodereasoning} evaluates input–output prediction for Python functions, CodeMMLU~\cite{manh2025codemmlumultitaskbenchmarkassessing} measures code comprehension across multiple languages and domains, and CRQBench~\cite{dinella2024crqbenchbenchmarkcodereasoning} derives reasoning questions from real-world code reviews. There are also more MLLM benchmarks contributed to various domains\cite{talmor2021multimodalqacomplexquestionanswering,zou2025dynamathdynamicvisualbenchmark,chow2025physbenchbenchmarkingenhancingvisionlanguage,NEURIPS2024_f2b9e8e7,NEURIPS2024_36b31e1b}. Furthermore, EMMA BENCH~\cite{hao2025mllmsreasonmultimodalityemma} (Enhanced MultiModal reAsoning) targets organic multimodal reasoning across mathematics, physics, chemistry, and coding. EMMA tasks require advanced cross-modal reasoning that cannot be solved by considering each modality independently, thus providing a challenging test suite for MLLMs.

\subsection{Material Science Benchmarks}
 LLM4Mat-Bench~\cite{rubungo2024llm4mat} is presented as the largest benchmark to date for evaluating the performance of large language models (LLMs) in predicting the properties of crystalline materials. The benchmark includes a vast dataset of approximately 1.9 million crystal structures from ten public databases, covering 45 distinct material properties. 
 ALDbench~\cite{yanguasgil2024benchmarkinglargelanguagemodels}, a new benchmark specifically designed to assess the capabilities of large language models (LLMs) in the domain of materials synthesis, consists of 70 open-ended questions, ranging from graduate to expert level, covering various aspects of ALD. 
MicroVQA~\cite{burgess2025microvqa} introduce a visual question answering benchmark in the field of biological microscopy. The benchmark is designed to evaluate three key reasoning capabilities crucial for scientific research: expert image understanding, hypothesis generation, and experiment proposal.
While current efforts have introduced benchmarks evaluating large language models for materials property prediction and synthesis, such as atomic layer deposition, there remains a lack of benchmarks that assess expert-level reasoning across diverse materials science domains, particularly those specializing in structure-property performance reasoning.
\begin{figure}[t]
  \centering
  \setlength{\belowcaptionskip}{-10pt}
  \includegraphics[width=1\textwidth]{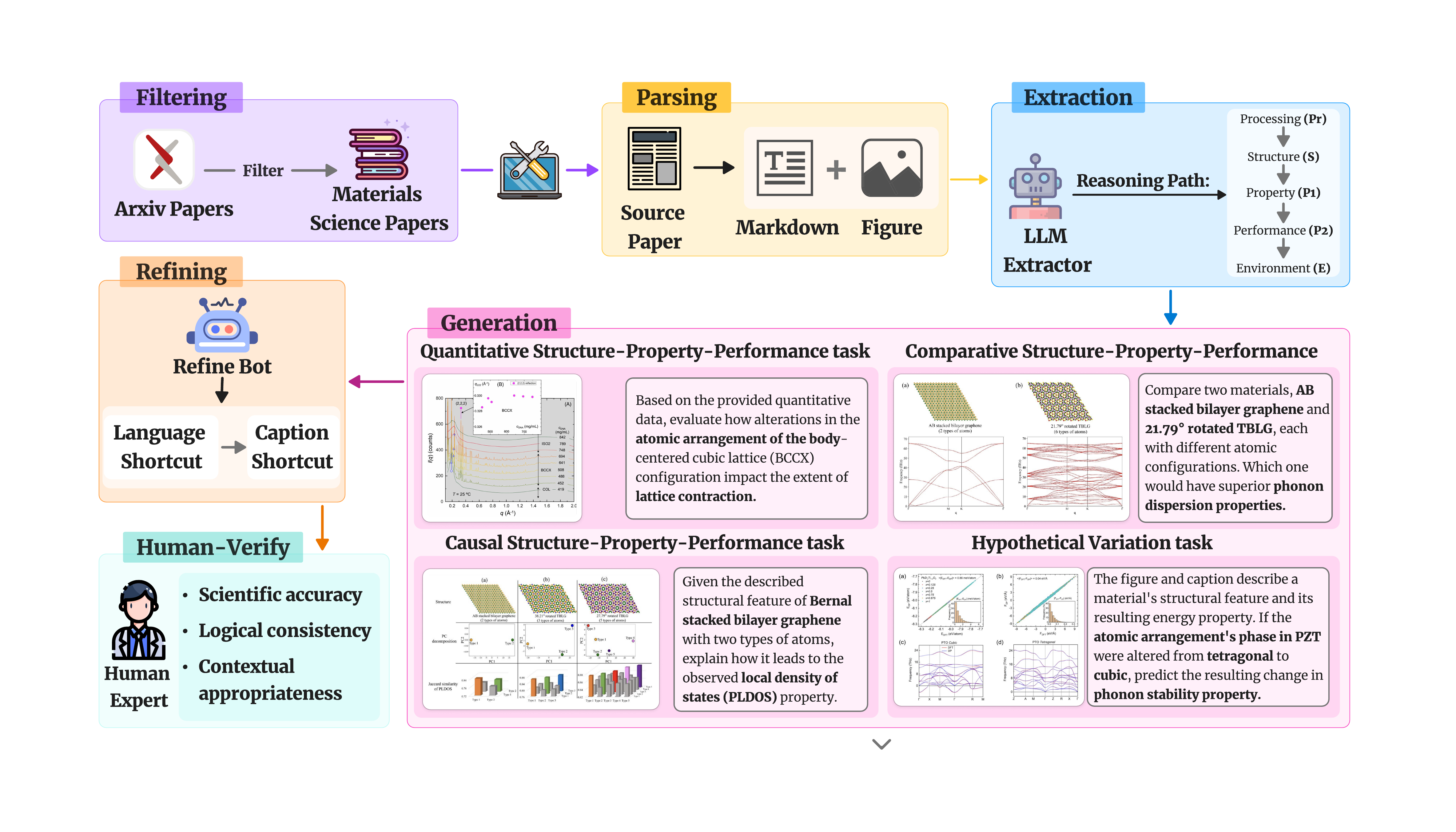}
  \caption{Construction Pipeline of MatVQA}
  \label{fig:main_construction}
  \vspace{-0.5em}
\end{figure}
\section{The MatVQA benchmark}
\subsection{Overview of MatVQA Benchmark}
 MatVQA is a fully synthetic dataset of 1325 VQA triplets.
The questions cover a broad spectrum in material science, ranging from metals, ceramics, electronic materials to coating. Fig~\ref{fig:domain-ratio} shows the distribution of MatVQA samples.  Especially, MatVQA closes three critical gaps in MLLMs in material science research:
\begin{enumerate}[label=(\arabic*),nosep]
\item \textbf{Domain coverage.} Existing MLLM benchmarks omit the complex, research-grade tasks that characterise materials science. MatVQA therefore formalises four \emph{structure–property–performance} (SPP) tasks that span microstructural analysis, mechanistic interpretation, and forward design—capabilities central to contemporary experimentation and modelling.
\item \textbf{Scientific updated.} Unlike datasets based on legacy exams, MatVQA draws its questions from the latest arXiv manuscripts, ensuring topical relevance and cutting-edge difficulty.
\item \textbf{Scalability with quality.} The \textbf{MArxivAgent} pipeline automatically extracts reasoning chains, generates high-fidelity MCQs, and removes language and caption shortcuts. A random 20\% of items are then audited by domain experts to guarantee research-grade accuracy.
\end{enumerate}
These innovations create the first benchmark capable of measuring an MLLM’s aptitude for precise, fine-grained visual reasoning across the full breadth of materials-science inquiry.
\subsection{Scientific Reasoning Tasks in MatVQA}
As depicted in Figure~\ref{fig:main_construction}, we design four high-impact tasks around five core components in material science—Structure (S), Property (P), Performance (Pe), Processing (Pr), and Environment (E):
\begin{description}[style=unboxed,leftmargin=0cm,nosep]
\item[Quantitative SPP.] Quantifies how structural parameters (e.g., grain size, defect density) affect performance metrics such as strength or ductility.
\item[Comparative SPP.] Compares materials with distinct architectures (e.g., framework vs.\ layered) to elucidate trade-offs in stiffness, toughness, and transport.
\item[Causal SPP.] Traces processing–structure–property causality, linking steps like dopant diffusion to device-level reliability.
\item[Hypothetical Variation.] Conducts “what-if” analyses of unrealised structural variants, guiding exploration of topological insulators, superconductors, and metamaterials.
\end{description}

Covering structural, energy, electronic, photonic, separation, catalytic, and environmental applications, these tasks integrate data science, high-throughput computation, and advanced characterization. Developed through expert interviews, representative examples are provided in the Appendix.
\begin{figure}[t]
    \centering
    \setlength{\belowcaptionskip}{-10pt}
    \includegraphics[width=1\linewidth]{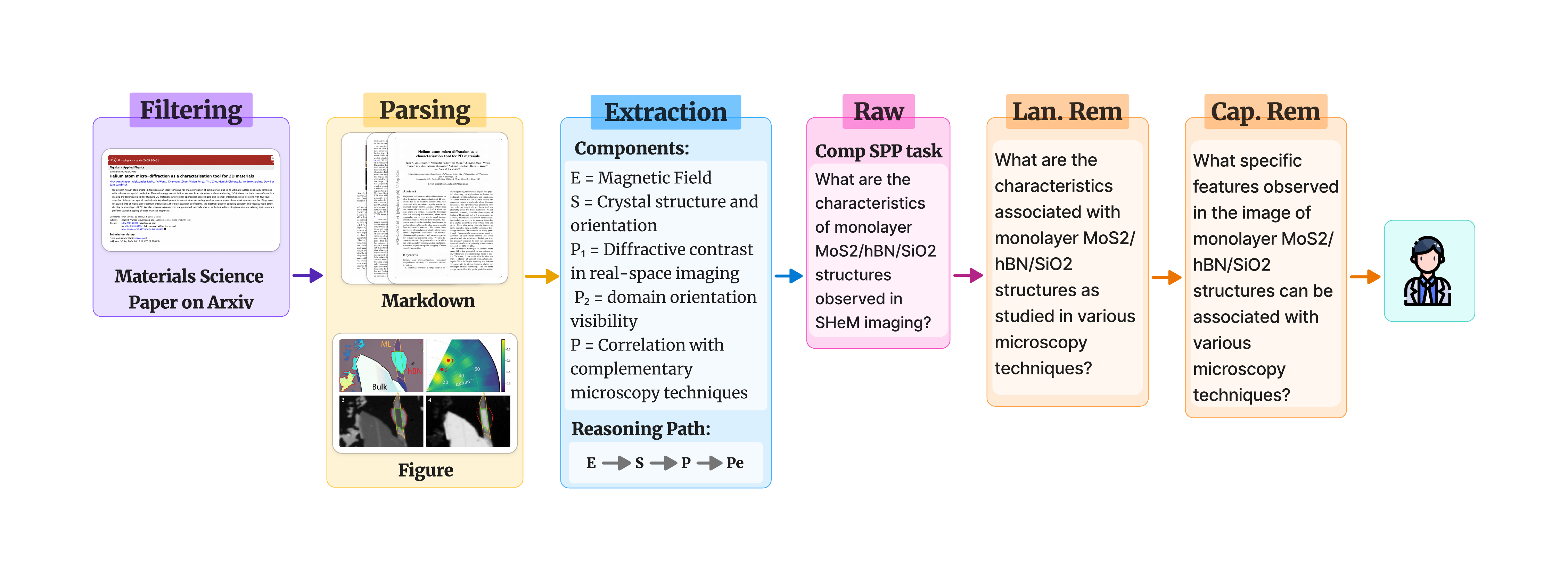}
    \caption{MArxivAgent Pipeline for MCQ automatically Generation. "Lan. Rem" represents the question after langauge shortcut removal. "Cap. Rem" represents the question after removing caption.}
    \label{fig:marxivagent}
    \vspace{-0.5em}
\end{figure}
\section{MArxivAgent: MatVQA Data-Generation Process}
This section details the pipeline used to create multiple-choice VQA items for MatVQA. As shown in Figure~\ref{fig:marxivagent}, the workflow has two main phases: (i) automated generation of raw visual question–answer triplets with verifiable reasoning chains and (ii) iterative shortcut removal followed by expert validation.
Creating VQA datasets is traditionally labour-intensive, for example, MicroVQA required 12 specialists and more than 600h. To eliminate this bottleneck we developed \textbf{MArxivAgent}, a fully automated system that yields research-grade, reasoning-centric MCQs for materials science.

\subsection{Phase 1: Generation of Raw VQA}
\textbf{Data Source.} MatVQA builds upon the arXiv open‑access platform\footnote{https://arxiv.org/}, leveraging granted remark and republication rights. We first retrieved 500 materials‐science articles published on 2024 via the arXiv API. Next, we filtered this corpus using task‑related keywords such as "property", "structure",etc. Because most manuscripts lack accessible source files, we employed Marker\footnote{https://github.com/VikParuchuri/marker} to extract figures and captions directly from the PDF collection. Beyond these visual elements, we also harvested the surrounding textual context for each figure—capturing extended explanations and analyses critical for complex reasoning. 

\textbf{Verifiable Reasoning Path Extractor.} To generate questions that probe complex reasoning over scientific figures, MArxivAgent first derives the most comprehensive reasoning chain from the provided caption and its surrounding context. This chain is structured around five core components—Structure (S), Property (P), Performance (Pe), Processing (Pr), and Environment (E)—with an emphasis on modeling their interrelationships. To keep consistencey identification of these elements, we incorporate the MatOnto\footnote{https://matportal.org/ontologies/MATONTO} in the components identification step. To ensure the extracted reasoning chains align with the paper, we further verify the confidence of each reasoning step by further retrieving the evidence in the original paper text, which makes the generated reasoning chain verifiable.

\textbf{MCQ construction.} Given a figure, its caption, and the verified reasoning path, the agent generates task-specific questions—causal, hypothetical, comparative, or quantitative—and rewrites them in MCQ format. 

\begin{wrapfigure}{r}{0.5\textwidth}
  \vspace{-3em}
    \begin{minipage}{0.5\textwidth}
      \begin{algorithm}[H]
        \caption{Two-stage Refinement}
        \label{alg:two_stage_refine}
        \begin{algorithmic}[1]
          \Require Raw MCQ set $Q=\{(q_i, C_i)\}$, where each \( C_i \) comprises the correct answer $A_i $ and distractors $D_i$, maximum iteration count $T$
          \Ensure A refined MCQ set $Q'$
          \State $Q' \leftarrow Q$
          \Statex \textbf{Stage 1: Language Shortcut Removal}
          \For{each $(q,A,D)$ in $Q'$}
            \State $t\leftarrow0$
            \While{$t<T$}
              \State $r\leftarrow \text{Evaluator}(q,C)$
              \If{not $r.\mathrm{correct}$} \State \textbf{break} \EndIf
              \State $s\leftarrow \text{ExtractLangStrategy}(r)$
              \State $(q^{'},C^{'})\leftarrow \text{Rewriter}(q,C,s)$
              \State $v\leftarrow \text{Checker}(q,A;q^{'},A^{'})$
              \If{ $\mathrm{semantically\_inconsistent}$}
                \State FAIL\_REWRITE and \textbf{continue}
            \EndIf
              \State $t\leftarrow t+1$
            \EndWhile
          \EndFor
          \Statex \textbf{Stage 2: Caption Shortcut Removal}
          \For{each $(q,C)$ in $Q'$}
            \State $t\leftarrow0$
            \While{$t<T$}
              \State $r\leftarrow \text{Evaluator}(q,C,captions)$
              \If{not $r.\mathrm{correct}$} \State \textbf{break} \EndIf
              \State $s\leftarrow \text{Reflector}(r)$
              \State $(q^{'},C^{'})\leftarrow \text{Rewriter}(q,C,s)$
              \State $v\leftarrow \text{Checker\_{caption}}(q,A;q^{'},A^{'})$
              \If{$\mathrm{reasoning\_path\_inconsistent}$}
                \State FAIL\_REWRITE and \textbf{continue}
                \EndIf
              \State $t\leftarrow t+1$
            \EndWhile
          \EndFor
          \State \Return $Q'$
        \end{algorithmic}
      \end{algorithm}
    \end{minipage}
    \vspace{-4.5em}
\end{wrapfigure}
Overall, this stage produced 1325 multiple-choice items: 950 causal, 256 hypothetical, 112 comparative, and just 7 quantitative. The pronounced imbalance arises from our reasoning-path constraint: every path must culminate in a performance node, a requirement that is rarely satisfied by figures with explicit numerical SPP data and therefore suppresses the generation of quantitative questions.

\subsection{Phase 2: Two-Stage Refinement}
The generated raw MCQs is inadequate for testing MLLM's capabilities. On one hand, the distractors are easily eliminated based on general material knowledge or too vague compared to correct option. Language shortcuts-information in the MCQ that allows answering the question without access to the image as proved in ~\cite{burgess2025microvqa}. On the other hand, we found that some questions can be directly answered by the figure caption without access to the figure. Complex reasoning required perception on the given image and further do multi-hop reasoning on the provided information. While the figure caption can only describe the main idea of this figure rather than demonstrate every details of the figure. We defined this kind of shortcut as caption shortcut. Therefore, we aim to construct questions removing these two shortcut to have high-quality MCQ distractors and questions required both fine-grained perception and multi-hop reasoning. The overall two‐stage refinement procedure is summarized in Algorithm~\ref{alg:two_stage_refine}.
 
 \textbf{Stage1: Language shortcut removal}
Here we follow the method in MicroVQA to remove the language shortcut to increase MCQ complexity: first apply an evaluator agent to answer the MCQ without image and then summary the COT answer as langauge strategy which will be passed to the rewriter LLM agent. The rewriter agent revises the original question and distractors to invalidate the language strategies. To aviod the significant changes on the revised question-answering pairs, an LLM checker is applied to ensure semantic equivalent with the original pair. This process will iterate if the question can still be answered correctly without image or after the pre-setting maximum iteration times. In this way, we can increase the difficulty of these MCQs without changing the scientific problem in original question-answering pairs.

 \textbf{Stage2: Caption shortcut removal}
After removing language shortcut in Stage 1, the questions still lack of fine-grained perception on figures since questions can be answered directly by the caption. Therefore, we introduce second stage refining MCQs by removing the caption shortcut. Similarly, we first use a evaluator agent to identify the caption shortcuts by answering the MCQs with caption only. And then reflecting on the strategies answering the question with a reflector LLM agent. The reflection results will further been passed to a rewriter to revise the question and options. Different from Stage 1, we only require the revised question-answering pairs follow the same reasoning chain used in Section 4.1, which results in larger modification on generated questions and aim to create harder problem removing the simple understanding pattern from captions.

An example of a generated reasoning chain is illustrated in Figure~\ref{fig:marxivagent}. In this sample case, provided with a paper from Arxiv. After parsing the paper pdf to markdown text and figures, we generate the reasoning path as: "10 T magnetic field (E) $\rightarrow$ collapse of the split magnetic Bragg satellites into a single peak (S) $\rightarrow$ suppression of the spin-cycloid and emergence of a collinear antiferromagnetic state (P) $\rightarrow$ redistribution of the total magnetic scattering into one commensurate reflection, doubling the peak intensity (Pe)." And for each reasoning step, we further verify it to grounded by the paper, which ensure the reliability of generated reasoning path.
\subsection{Human Expert Quality Check}
Because MatVQA is machine-generated end-to-end, a human audit was imperative. Limited by annotation resources, two materials-science experts independently reviewed a random 20\% sample of the MCQs, scoring each for scientific accuracy, logical consistency, and contextual relevance. The uniformly high marks in this audit gave us sufficient confidence to freeze the dataset and release the present benchmark version.

\section{Experiments}
\subsection{Benchmarking MLLMs with MatVQA}
To comprehensively evaluate the validity and difficulty of our proposed dataset, MatVQA, we conducted a systematic series of experiments across a range of state‑of‑the‑art Multimodal Large Language Models (MLLMs) as shown in Table~\ref{tab:vlm-performance}. The selected models encompass both open‑source and proprietary systems and represent leading capabilities in visual‑and‑language understanding, including Grok-2-Vision\cite{xai2024grok2vision} and LLaVA\cite{meta2024llama3}, as well as the closed‑source GPT‑4o\cite{gpt4o-mini} and Gemini\cite{google2025gemini2.5proexp}.  We utilize standard chain-of-thought prompting\cite{yue2024mmmu}(deatails in
Appendix). By comparing these models’ performance on the MatVQA benchmark, we aim to elucidate the current strengths, limitations, and avenues for improvement in multimodal material visual-reasoning question answering problems.

\begin{table*}[t]
  \centering
  \caption{Performance of various vision–language models on MatVQA by task: Causal Structure-Property-Performance task(Caus),Comparative Structure-Property-Performance task(Comp), Quantitative Structure-Property-Performance task(Quan),Hypothetical Variation task(Hypo).The evaluated models are splited to four parts: Reasoning mode, Large models, Small models and Material-finetuned model(Material-FT Models).}
  \label{tab:vlm-performance}
  \begin{tabular}{@{} >{\centering\arraybackslash}p{1.5cm} l c c c c c c @{}}
    \toprule
    \textbf{Category} 
      & \textbf{Model} 
      & \textbf{Overall} 
      & \textbf{Caus} 
      & \textbf{Hypo} 
      & \textbf{Quan}
      & \textbf{Comp}\\
    \midrule
    \multirow{1}{1.5cm}{\small\textbf{Reasoning}}
      & o1\cite{OpenAI2024} 
      & 48.6\% & 49.7\% & \textbf{48.0}\% & \textbf{71.4}\% & 39.3\% \\
    \midrule
    \multirow{9}{1.5cm}{\small\textbf{Large Models}}
      & Claude‑3.7‑Sonnet\cite{anthropic2025claude3.7sonnet}  &\textbf{51.9}\%  &\textbf{53.1}\% & 42.5\% & 57.1\% &  48.2\% \\
      & Gemini-1.5-Pro\cite{geminiteam2024gemini1p5}     & 44.2\% & 44.8\% & 41.5\% & 57.1\%  &  43.7\%\\
      & Mistral-small-3.1-24b-instruct\cite{mistralai2025mistralsmall3.1}       & 45.7\% & 45.9\% & 44.9\% & 57.1\% & 44.6\%\\
      & Grok‑2‑Vision\cite{xai2024grok2vision}      & 47.4\%  & 48.7\% & 43.4\% & \textbf{71.4}\% & 43.7\%\\
      & Qwen‑2.5‑vl‑72b‑Instruct\cite{QwenTeam2025} & 44.7\% & 46.0\% & 39.8\% & 57.1\%  & 41.9\%\\
      & GPT‑4o\cite{gpt4o-mini}              & 48.3\% & 50.0\% & 42.5\% &  \textbf{71.4}\% & 45.5\%\\
      & Llama‑3.2‑90b‑Vision‑Instruct\cite{meta2024llama3}   & 39.1\%& 40.0\% & 32.8\% & 28.5\% & 46.4\% \\
      & Phi-4-Multimodal-Instruct\cite{microsoft2025phi4multimodal}   & 28.2\% & 27.2\% & 30.1\%  & 57.1\% & 27.6\%\\
    \midrule

    \multirow{7}{1.5cm}{\small\textbf{Small Models}}
      & Qwen‑2.5‑VL‑7b-Instruct\cite{QwenTeam2025}       &42.6\%  & 44.0\% & 38.7\%  & 28.5\% & 40.1\% \\
      & Claude‑3.5‑Haiku\cite{anthropic2024claude3.5haiku}    & 44.7\% & 32.9\% & 38.3\% & 57.1\% & 37.5\% \\
      & Gemini‑Flash‑2.0\cite{google2025geminiflash2.0} & 44.8\%  & 44.4\% & 42.9\% & 28.5\% & \textbf{53.5}\%  \\
      & GPT‑4o‑mini\cite{gpt4o-mini}         & 39.3\% & 41.1\% & 34.3\%  & 71.4\% &  33.0\%\\
      & Pixtral‑12b\cite{clarifai2024pixtral12b}          & 41.5\%  &43.8\% & 35.2\%  & 42.8\% &  36.6\%\\
      & Llama‑3.2‑11b‑vision‑instruct\cite{meta2024llama3}  &32.5\% & 32.9\% & 28.1\% & 57.1\%& 36.6\% \\
    \midrule
    \multirow{2}{1.5cm}{\small\textbf{Material-FT Models}}
    & MOL-VL-7B\cite{luo2023biomedgpt}  & 23.6\% & 32.9\% & 22.7\% & 44.7\% & 2.8\% \\
    & Cephalo-8B-Alpha\cite{Buehler_Cephalo_2024} & 23.8\% & 22.4\% & 25.4\% & 28.6\% & 31.4\% \\
    \bottomrule
  \end{tabular}
  \vspace{-0.5em}
\end{table*}

\textbf{MatVQA is uniformly challenging} – The highest overall performanced  is achieved by \textsc{Claude-3.7-Sonnet}(51.9\%) from the "large models" category. Reasoning model O1 shows a strong baseline performance at 48.6\%. In the small model category, \textsc{Gemini-Flash-2.0} (44.8\%) and Claude-3.5-Haiku (44.7\%) were the top performers, indicating that smaller architectures can achieve competitive results with large models. The domain-specific model \textsc{MOL-VL-7B} shows the lowest overall performance across all categories at 23.6\%. The reason might related that MOL model is finetuned for optical chemical structure understanding which bias the output for our data. The uniformly low accuracy proved that MatVQA is challenging for both large language models and small language models. These limitations likely stem from a combination of factors, including the nuanced visual perception required for material-scientific figures and the sophisticated reasoning demanded by tasks such as comparative and hypothetical analysis, which were identified as particularly challenging.

\textbf{Scale helps, but only up to a point} – As shown in Table \ref{tab:vlm-performance}, the large-parameter models on average surpass their small-parameter counterparts by +4.6 pp on the dominant Causal split (44.5\% vs. 39.9\%) and by +9.5 pp on the Quantitative split (57.1\% vs. 47.6\%). Yet sheer size is no guarantee of success: the 90 B \textsc{Llama-3.2-90b-Vision} (39.1\%) and 72 B \textsc{Phi-4-Multimodal} (28.2\%) lag well behind the 12 B \textsc{Gemini-Flash-2.0} (44.8\%) and even the 7 B \textsc{Claude-3.5-Haiku} (44.7\%). These outliers illustrate that clever architectural design and task-specific multimodal fine-tuning can outweigh brute-force scaling. In short, bigger models generally deliver a performance premium on MatVQA, but the marginal gains taper off and vary sharply by task. A likely reason is that MatVQA requires fine-grained, domain-specific cross-modal reasoning about materials, which depends less on parameter count than on how effectively the vision and language components are aligned and how much relevant scientific data the model has actually seen.

\textbf{Difficulty varies strongly by task} – Averaging across size tiers, large-parameter models score 44.5\% on the Causal split, 39.7\% on Hypothetical, 42.7\% on Comparative, and 57.1\% on the sparsely populated Quantitative split, while small models reach 39.9\%, 36.3\%, 39.6\%, and 47.6\% respectively. Causal reasoning is both the most fundamental skill for materials science and the backbone of the benchmark (950 of 1325 items), so modest gains here translate directly into meaningful real-world impact. Comparative questions, though not as numerous, record the worst performance; they demand precise joint perception of two entities and consistency across several reasoning hops, exposing brittleness in current vision–language pipelines. By contrast, Quantitative items appear “easy,” yet they comprise only seven test cases, so headline scores primarily reflect the tougher causal and comparative challenges rather than a few numeric pattern-matching problems.

These results collectively demonstrate that MatVQA provides a stringent and well-balanced yard-stick: it highlights genuine progress in multimodal scientific QA while exposing persistent weaknesses—particularly in counter-factual design reasoning—that future MLLMs must overcome.

\subsection{Ablation for Two-Stage Refinement}
\begin{figure}[t]
  \centering
  \includegraphics[width=1\textwidth]{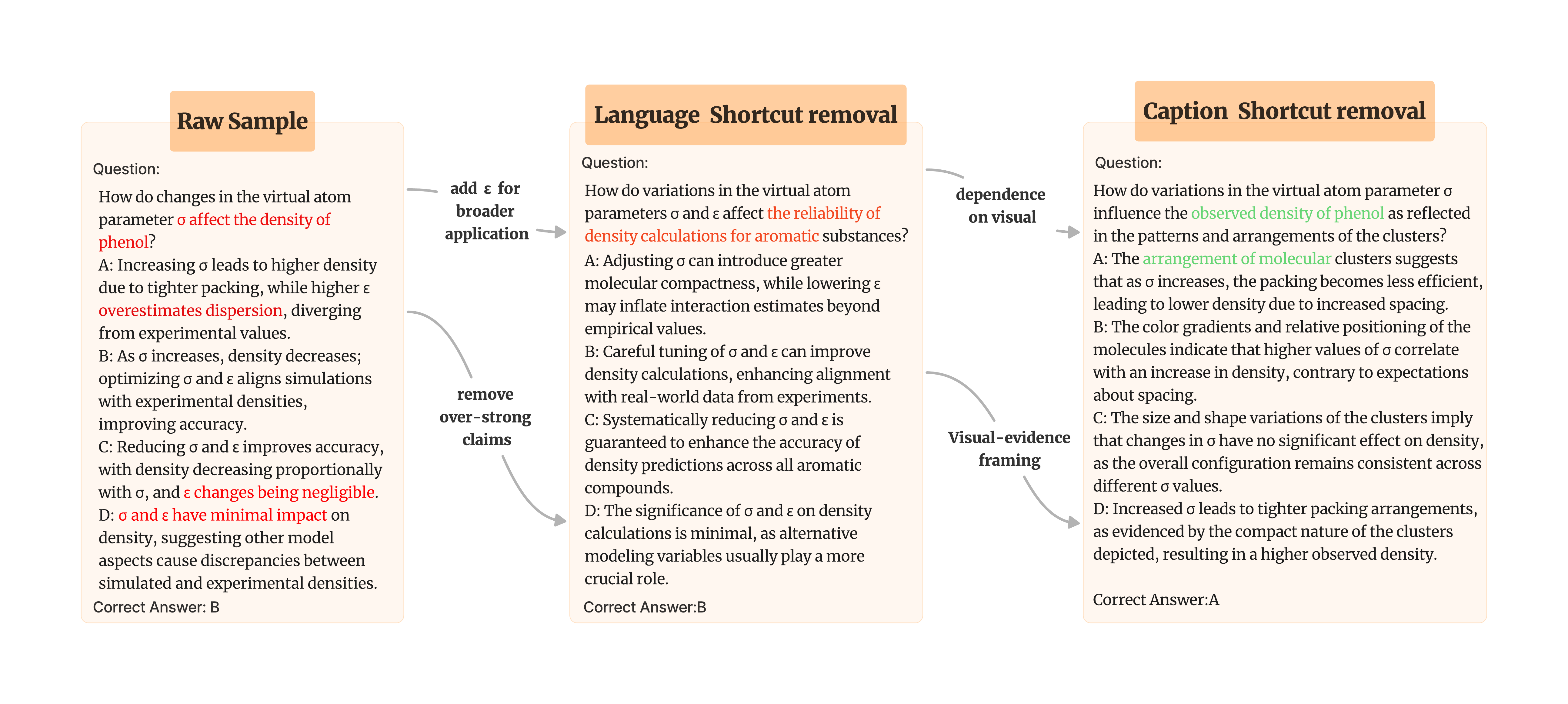}
  \caption{Evolution of a sample question through the two-stage shortcut removal process. The figure shows the transformation from:  the initial 'Raw Sample,' to after 'Language Shortcut removal', and finally to after 'Caption Shortcut removal'.}
  \label{fig:exemplar}
  \vspace{-0.5em}
\end{figure}

\begin{table*}[t]
\caption{Two-stage Refinement evaluation results. "Lan.Rem" represents the results after removing language shortcut and "Cap.Rem" represents the results after removing caption shortcut.}
\scriptsize
\centering
\begin{tabular*}{\textwidth}{@{\extracolsep{\fill}}lcccccc@{}}
  \toprule
  \textbf{Stage} & \textbf{GPT-4o} & \textbf{GPT-4o-mini} & \textbf{Claude-3.7-Sonnet} & \textbf{Claude-3.5-Haiku} & \textbf{o1} & \textbf{Gemini-1.5-Pro} \\
  \midrule
  Raw     & 78.8\% & 72.4\% & 82.7\% & 75.3\% & 79.6\% & 76.4\% \\
  Lan.Rem & 67.8\% (\textbf{11.0}\%$\downarrow$) & 57.4\% (\textbf{15.0}\%$\downarrow$) & 71.6\% (\textbf{11.1}\%$\downarrow$) & 65.0\% (\textbf{10.3}\%$\downarrow$) & 68.3\% (\textbf{11.3}\%$\downarrow$) & 63.2\% (\textbf{13.2}\%$\downarrow$) \\
  Cap.Rem & 48.3\% (\textbf{30.5}\%$\downarrow$) & 39.3\% (\textbf{33.1}\%$\downarrow$) & 51.9\% (\textbf{30.8}\%$\downarrow$) & 44.7\% (\textbf{30.6}\%$\downarrow$) & 48.6\% (\textbf{31.0}\%$\downarrow$) & 44.2\% (\textbf{32.2}\%$\downarrow$) \\
  \bottomrule
\end{tabular*}
\label{tab:two_stage_drop}
\vspace{-1em}
\end{table*}
The results in Table~\ref{tab:two_stage_drop} demonstrate that both language and caption refinement stages significantly reduce model accuracy, compelling a deeper level of reasoning. Language shortcut removal (Stage 1, or Lan.Rem) decreases accuracy by approximately 12\% on average (individual model drops typically range 10\%–15\%). As illustrated by the sample question evolution in Figure~\ref{fig:exemplar}, this initial difficulty increase stems from broadening the question's scope from specific details (e.g., ‘phenol’) to more generalized concepts (e.g., ‘aromatic substances’) and removing overly assertive or simplistic textual cues. This demands more nuanced language comprehension from the models. Scientifically, this stage appropriately challenges models, as overgeneralizing concepts without full contextual understanding can lead to misinterpretations in many real-world phenomena. 

Caption shortcut removal instigates a more substantial performance decline, with an additional average accuracy drop of approximately 19\% from language removal (individual model drops typically range 18\%–20\%). By compelling models to value low-level visual perception from experimental figures—such as the "patterns and arrangements of the clusters" shown in the example—instead of relying on textual hints, Caption removal ensures a rigorous test of genuine visual grounding and fine-grained perception. This is a crucial step towards evaluating true multimodal understanding.

\subsection{Error Analysis}
To further explore the current shortcomings of MLLMs, we further analysis the CoT responses of error samples. We further classify the errors into three types.

\textbf{Visual Perception Error}- the model lifts the wrong information from the figure itself (e.g., reversing a plotted trend, overlooking obvious irregular gaps, or confusing one image panel with another), so every later step starts from an incorrect visual cue.

\textbf{Material Knowledge Misunderstanding} — the chain-of-thought contains a coherent deduction, but it rests on a faulty scientific premise (e.g., assuming larger intermolecular spacing does not weaken binding, or believing broader peaks imply sharper selectivity), leading to a confident yet wrong choice.

\textbf{Reasoning Wrong Judgement} — the model’s logic falters even after perceiving the data correctly: it misreads option wording, evaluates the wrong option set, or applies an inconsistent elimination rule, so the final selection contradicts its own stated observations.


\section{Conclusion}
In this work, we presented MatVQA, a benchmark that marks a significant step towards evaluating true multimodal reasoning in materials science. MatVQA compels Multimodal Large Language Models (MLLMs) to genuinely "look" at complex material experimental figures, engaging with fine-grained visual details rather than relying on textual shortcuts. This is achieved through our fully automated MArxivAgent pipeline—a robust and adaptable methodology with strong potential for application across diverse scientific fields beyond materials science—which, for the current benchmark, rigorously eliminates language and caption artifacts from recent arXiv publications in materials science. This automated approach not only produces an initial set of 1325 high-fidelity questions grounded in visual evidence across four key structure-property-performance reasoning tasks (quantitative, comparative, causal, and hypothetical) but also ensures MatVQA’s inherent scalability.
Looking forward, we plan to leverage this scalability to expand MatVQA to approximately 12,000 questions. This substantial expansion will significantly enhance its comprehensiveness and utility in rigorously assessing MLLM capabilities. Such growth, potentially alongside future extensions into areas like 3D crystal structures, will be crucial for advancing MLLMs that can truly understand and reason about visually rich scientific data, thereby fostering their meaningful contribution to materials research.

\bibliographystyle{plain}
\bibliography{refer}

\newpage
\setcounter{page}{1}
\appendix

\section{Prompts For MArxivAgent}
All prompts for evaluation used the chain-of-thought prompt template from the MMMU-Pro code\cite{yue2024mmmu}:
\begin{figure}[h]
    \centering
    \includegraphics[width=\linewidth]{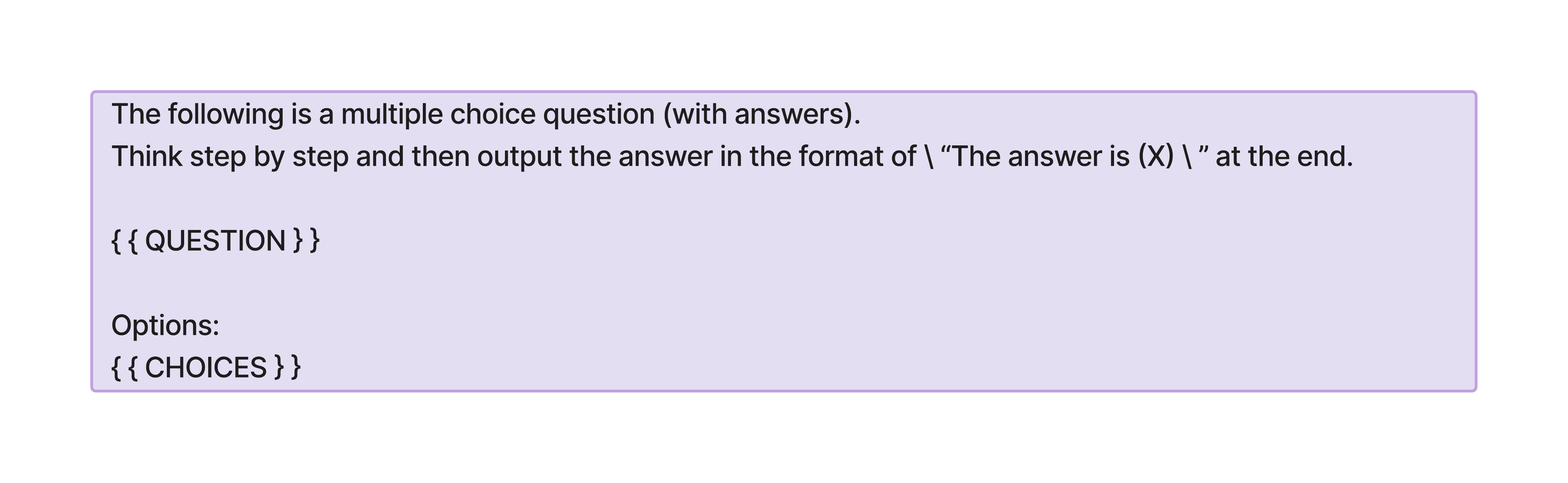}
    \label{fig:enter-label}
\end{figure}

An example complete question is:
\begin{figure}[h]
    \centering
    \includegraphics[width=\linewidth]{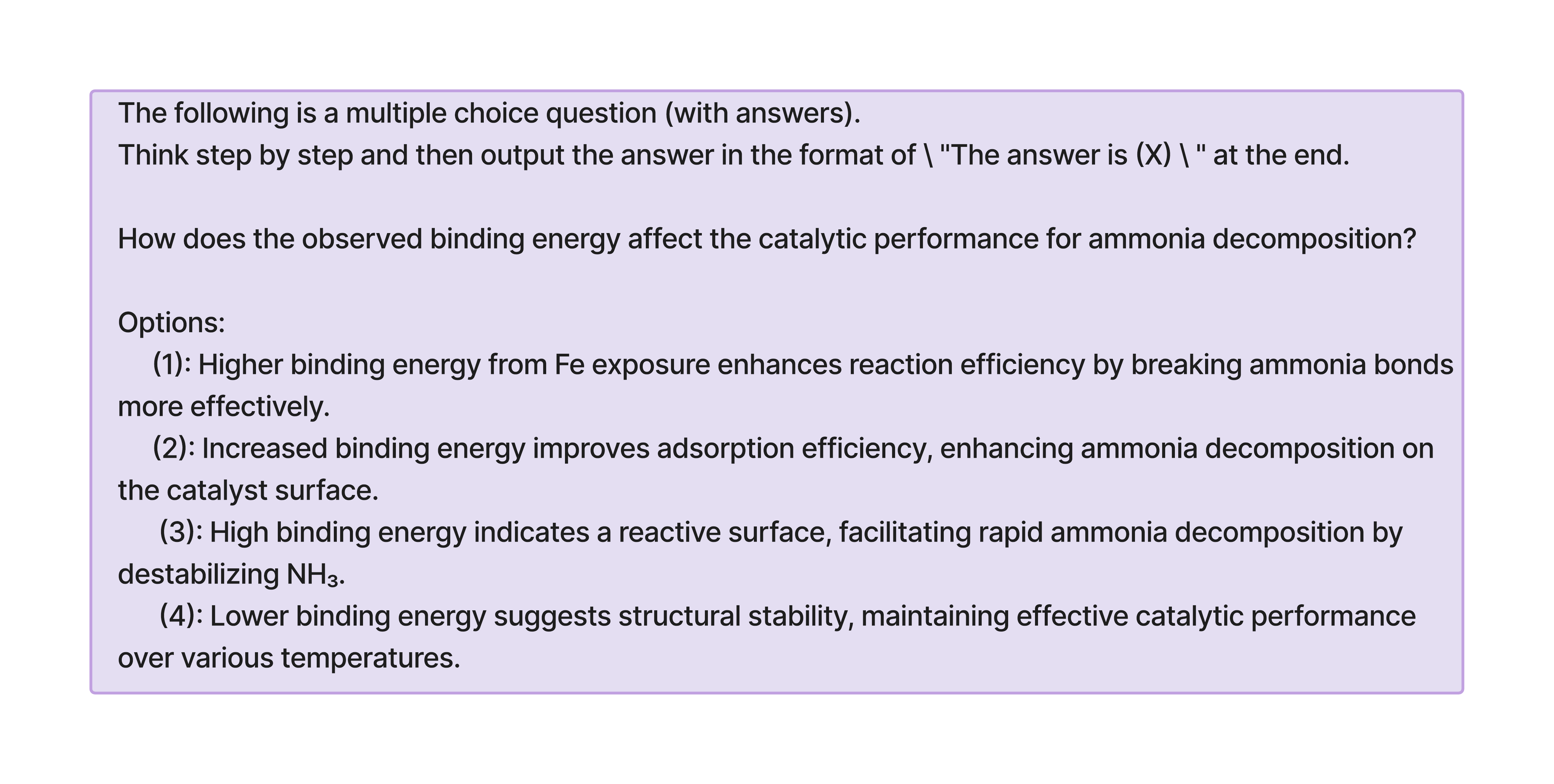}
    \label{fig:enter-label}
\end{figure}

Candidate answers are extracted with the regular expression\
\verb|answer is\s*([0-9])|.

\newpage
\section{Representative Examples}
\begin{figure}[h]
    \centering
    \includegraphics[width=0.8\linewidth]{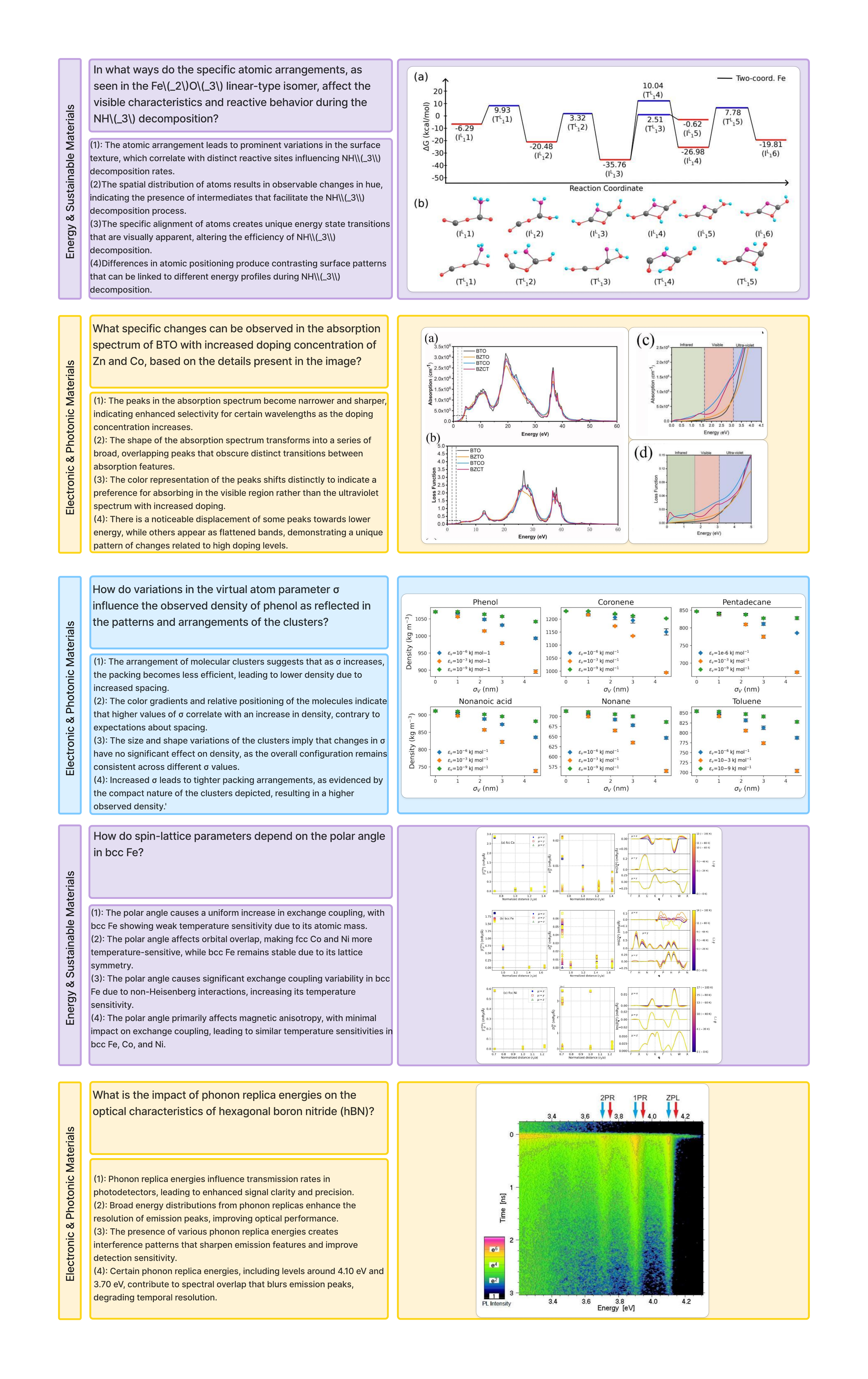}
    \caption{Representative Examples for varies material science domain.}
    \label{fig:enter-label}
\end{figure}

\newpage
\section{Failure Analysis}
\begin{figure}[h]
    \centering
    \includegraphics[width=1\linewidth]{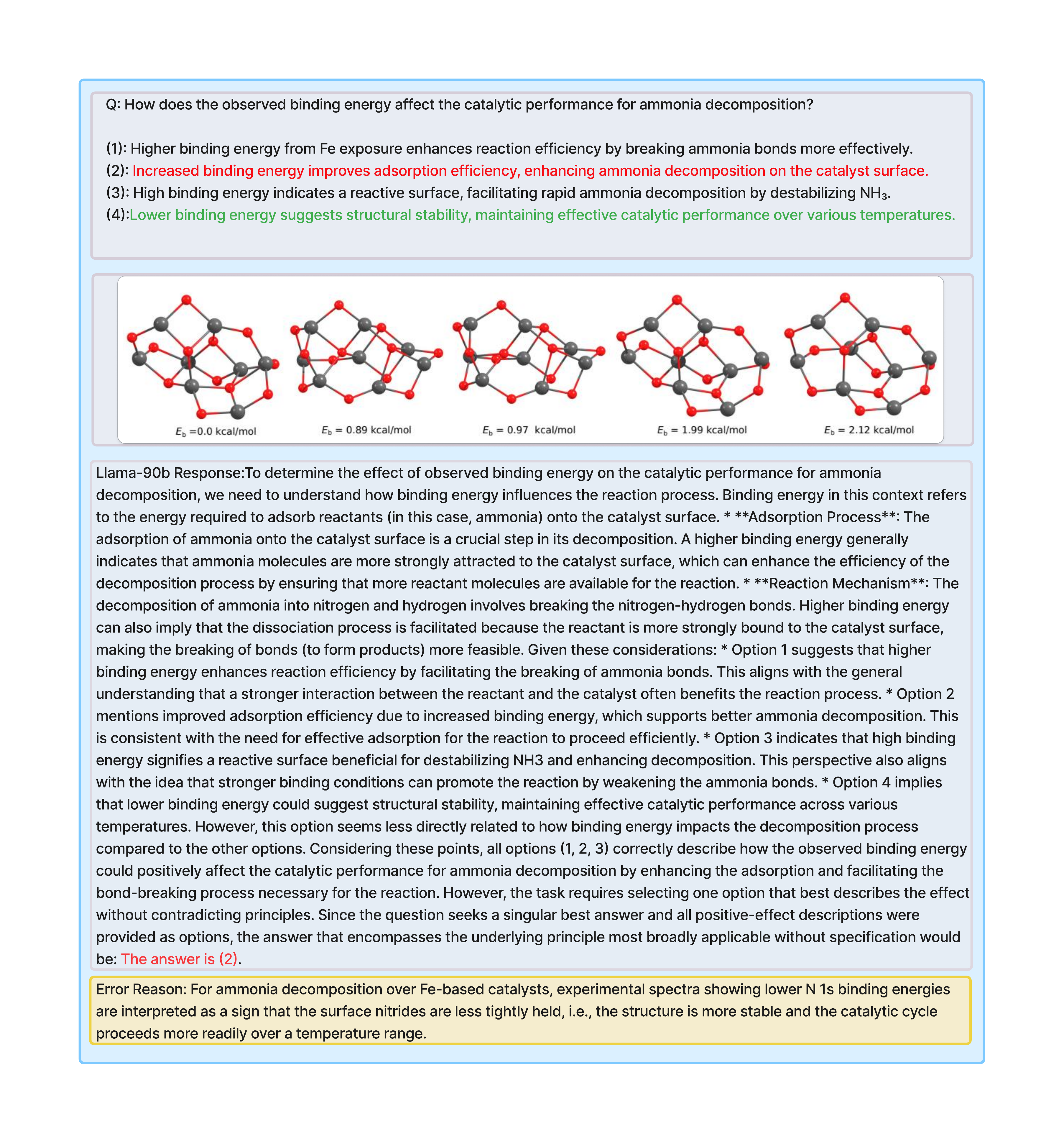}
    \caption{Error sample for Material Knowledge Misunderstanding error type, the response missed catalytic know-how—namely the Sabatier optimum and the distinction between Fe–support binding vs. $NH_{3}$ adsorption}
    \label{fig:error0}
\end{figure}

\begin{figure}[h]
    \centering
    \includegraphics[width=0.95\linewidth]{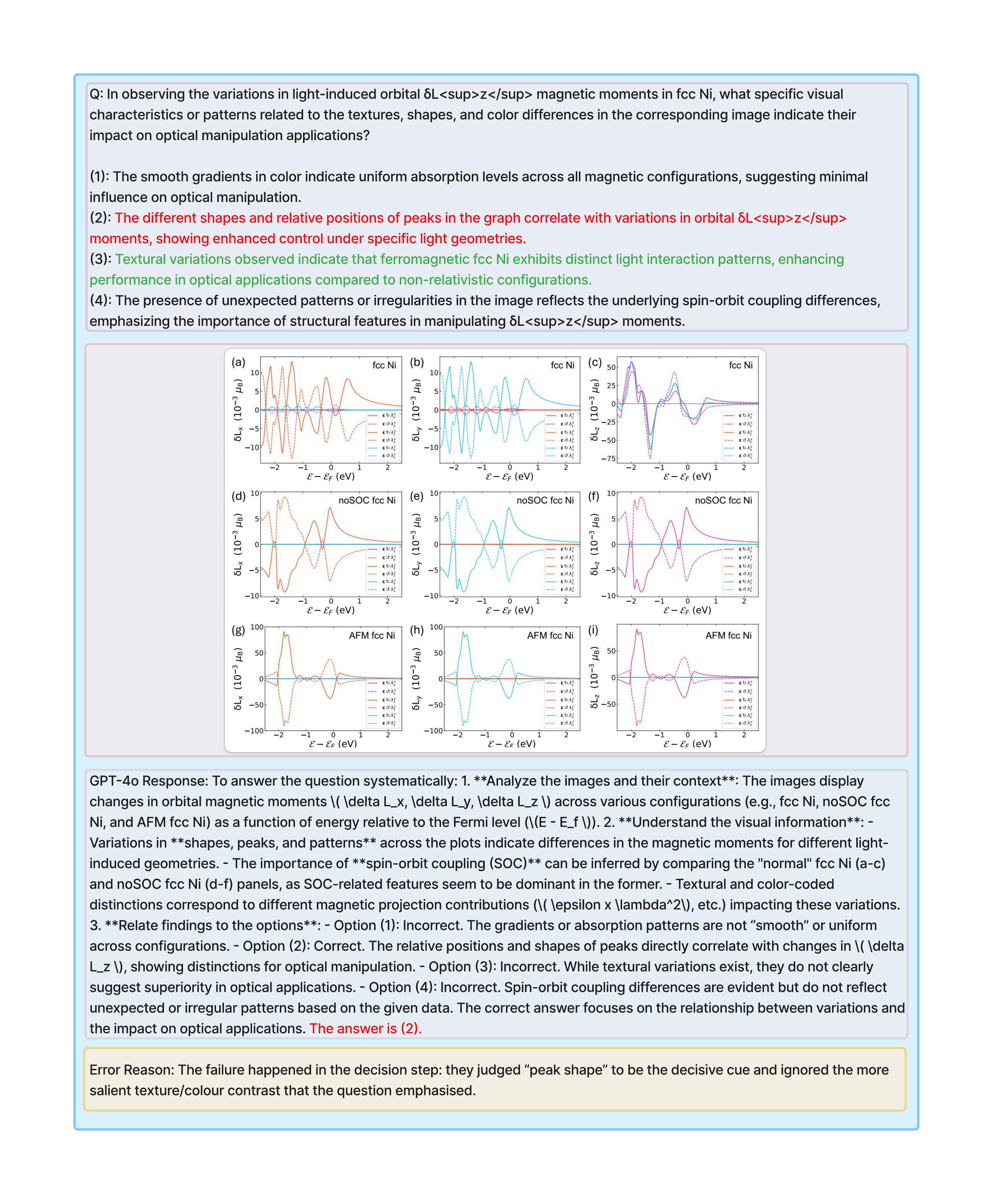}
    \caption{Error sample for Reasoning-Wrong-Judgement error type. In this example, optical control schemes exploit those extra SOC-enabled orbital channels; the richer the texture, the more knobs you have. Therefore the figure is telling us that ferromagnetic fcc Ni, with its unique textural signature, offers superior optical-manipulation capability compared with the non-relativistic case. The respondent saw the figure correctly (they talked about peaks and SOC). They knew material concepts (spin-orbit coupling, ferromagnetism).
The failure happened in the decision step: they judged “peak shape” to be the decisive cue and ignored the more salient texture/colour contrast that the question emphasised.}
    \label{fig:error1}
\end{figure}

 Several error examples are shown in Figure\ref{fig:error0} and Figure\ref{fig:error1}.

\newpage

\end{document}